\documentclass[conference, A4]{IEEEtran}
\IEEEoverridecommandlockouts

\usepackage{cite}
\usepackage{amsmath,amssymb,amsfonts}
\usepackage{algorithmic}
\usepackage{graphicx}
\usepackage{textcomp}
\usepackage{xcolor}
\usepackage[utf8]{inputenc}
\usepackage{hyperref}
\usepackage{cite}
\usepackage{xcolor}
\usepackage[normalem]{ulem}
\usepackage{float}
\usepackage{caption}
\usepackage{subcaption}
\usepackage{todonotes}
\usepackage[ruled, linesnumbered]{algorithm2e}
\usepackage{comment}
\usepackage{amsthm}

\usepackage{bm}
\def\BibTeX{{\rm B\kern-.05em{\sc i\kern-.025em b}\kern-.08em
    T\kern-.1667em\lower.7ex\hbox{E}\kern-.125emX}}
\begin{document}

\newtheorem{remark}{Remark}
\newtheorem{definition}{Definition}
\newtheorem{theorem}{Theorem}
\newtheorem{invariant}{Invariant}

\newcommand{\GF}[1]{\noindent\textcolor{brown}{{\fontfamily{phv}\selectfont GF-NOTE: #1}}}
\newcommand{\SB}[1]{\noindent\textcolor{purple}{{\fontfamily{phv}\selectfont SB-NOTE: #1}}}
\newcommand{\ST}[1]{\noindent\textcolor{blue}{{\fontfamily{phv}\selectfont ST-NOTE: #1}}}
\newcommand{\RF}[1]{\noindent\textcolor{olive}{{\fontfamily{phv}\selectfont RF-NOTE: #1}}} 

\title{A New Probabilistic Mobile Byzantine Failure Model for Self-Protecting Systems}

\makeatletter
\newcommand{\linebreakand}{%
  \end{@IEEEauthorhalign}
  \hfill\mbox{}\par
  \mbox{}\hfill\begin{@IEEEauthorhalign}
}
\makeatother


\author{\IEEEauthorblockN{Silvia Bonomi}
\IEEEauthorblockA{
\textit{Sapienza University of Rome}\\
Via Ariosto 25, 00185, Rome, Italy \\
bonomi@diag.uniroma1.it}
\and
\IEEEauthorblockN{Giovanni Farina}
\IEEEauthorblockA{
\textit{Niccol\'o Cusano University}\\
Rome, Italy \\
giovanni.farina@unicusano.it}
\and
\IEEEauthorblockN{Roy Friedman}
\IEEEauthorblockA{
\textit{Technion}\\
Israel\\
roy@technion.ac.il}
\linebreakand
\IEEEauthorblockN{Eviatar B. Procaccia}
\IEEEauthorblockA{
\textit{Technion}\\
Israel \\
eviatarp@technion.ac.il}
\and
\IEEEauthorblockN{Sebastien Tixeuil}
\IEEEauthorblockA{
\textit{Sorbonne University, CNRS, LIP6, IUF}\\
France\\
sebastien.tixeuil@lip6.fr}

}

\maketitle

\thispagestyle{plain}
\pagestyle{plain} 

\begin{abstract}
Modern distributed systems face growing security threats, as attackers continuously enhance their skills and vulnerabilities span across the entire system stack, from hardware to the application layer. 
In the system design phase, fault tolerance techniques can be employed to safeguard systems. 
From a theoretical perspective, an attacker attempting to compromise a system can be abstracted by considering the presence of Byzantine processes in the system. 
Although this approach enhances the resilience of the distributed system, it introduces certain limitations regarding the accuracy of the model in reflecting real-world scenarios.
In this paper, we consider a self-protecting distributed system based on the \emph{Monitoring-Analyse-Plan-Execute over a shared Knowledge} (MAPE-K) architecture, and we propose a new probabilistic Mobile Byzantine Failure (MBF) that can be plugged into the Analysis component. 
Our new model captures the dynamics of evolving attacks and can be used to drive the self-protection and reconfiguration strategy.
We analyze mathematically the time that it takes until the number of Byzantine nodes crosses given thresholds, or for the system to self-recover back into a safe state, depending on the rates of Byzantine infection spreading \emph{vs.} the rate of self-recovery.
We also provide simulation results that illustrate the behavior of the system under such assumptions.
\end{abstract}


\section{Introduction}
Modern applications make use of distributed systems to eliminate potential computational bottlenecks, improve performance, and ensure reliability. 
While distributed environments can help overcome certain failures, they also introduce higher security risks~\cite{anderson2020security,DBLP:conf/dsn/LiSTWL21}. 
Repairing the system in such scenarios can become a time- and resource-intensive process~\cite{DBLP:conf/dsn/ZhaoCBRBMC19}.
This led to increased attention toward self-* systems i.e., systems able to support \emph{Self-configuration} (i.e., automatic configuration of components)~\cite{DBLP:conf/dsn/MartinSDBF15,monperrus2018automatic}, \emph{Self-healing} (i.e., automated discovery, and correction of faults)~\cite{DBLP:conf/dsn/PozoRH21}, \emph{Self-optimization} (i.e., autonomous monitoring and control of resources to ensure optimal functioning for the defined requirements)~\cite{maimo2018self}, and \emph{Self-protection} (i.e., proactive identification and protection from arbitrary attacks)~\cite{10.1145/2555611,zhang2018self}.

In this paper, we focus \ifdefined\CONFversion \else specifically \fi on \emph{self-protecting} systems, a subclass of autonomous systems that can detect and mitigate security threats in real-time~\cite{10.1145/2555611}.
Similar to other self-* properties, self-protection enables the system to autonomously adapt to constantly changing environments with minimal human intervention, making it responsive, agile, and cost-effective.

According to the conceptual model FORMS~\cite{FORMS}, a self-protecting system is based on the design principle of \emph{separation of concerns}. From an abstract point of view, a self-protecting system is immersed in an environment (that could be both a physical and a software environment). 
It consists of a two-layer architecture: (i) a managed subsystem layer (also called \emph{monitored environment}) that comprises the application logic, and a managed subsystem layer on top of the previous subsystem (also called \emph{autonomic or protecting environment}) comprising the adaptation logic.

\ifdefined\CONFversion
\else
\begin{figure}[t]
    \centering
    \includegraphics[width=0.5\linewidth]{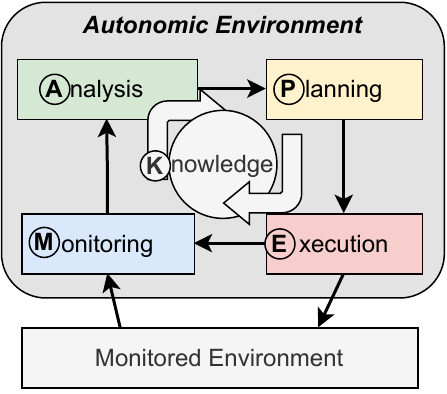}
    \caption{Monitor, Analyze, Plan, Execute over a shared Knowledge (MAPE-K) feedback loop phases.}
    \label{fig:mapek}
\end{figure}
\fi

While the design and implementation of the monitored environment are highly dependent on the context and on the specific application domain, the autonomic environment is typically organized in the form of a feedback control loop, leveraging the \emph{Monitor-Analyze-Plan-Execute over a shared Knowledge} (MAPE-K) architecture~\cite{1160055,arcaini2015mapek}.
\ifdefined\CONFversion
\else

The MAPE-K architecture consists of five conceptual modules:
\begin{itemize}
    \item a \emph{Monitor} component ({\bf M}) that collects data from the underlying monitored environment via probes (or sensors);
    \item an \emph{Analyze} ({\bf A}) component that performs data analysis to check whether an adaptation is required;
    \item a \emph{Plan} component ({\bf P}) that composes a workflow of adaptive actions necessary to achieve the system's objectives if required;
    \item an \emph{Execution} component ({\bf E}) that executes actions identified in the computed plan through actuators acting on the monitored system;
    \item a \emph{Knowledge Base} component ({\bf K}) that maintains data of the monitored environment, adaptation goals, and other relevant states that are shared by the MAPE components.
\end{itemize}


\fi
In this work, the monitored environment is represented by a distributed system made of a set of $n$ processes, collaborating to run a distributed application. Processes may get compromised by an external attacker who lets them behave arbitrarily (i.e., it makes them Byzantine faulty).
More in detail, we consider a mobile variant of the Byzantine failures model~\cite{ByzGen}, in which Byzantine failures may spread in the system with some probability or rate due to the progressive action of the attacker.
Similarly, each node individually, as well as the entire system, can trigger a recovery action, a process that is also known as \emph{rejuvenation}~\cite{SFV05,SNVS06}.
Such local recovery may occur at a given rate as well, or we may assume that an infected node may be able to detect its own intrusion with a given probability through a local trusted computing monitor, which would then spark a local recovery from a fresh safe~copy.

Naturally, local recovery is preferred, since it is relatively fast and does not interrupt the entire system's operation.
On the other hand, there may be situations in which failures are so spread that local recoveries are not likely to bring the system back to a safe state, and therefore a coordinated full system rejuvenation is needed.
Being able to make such predictions is important both from an operational standpoint, as well as for configuration and planning purposes, and in order to be able to make reliability and availability statements.
\smallskip

\noindent{\bf Our contribution.}
We discuss how to enrich a distributed system designed with a MAPE-K-based architecture that offers self-protection capabilities.
In particular, we propose a new probabilistic Mobile Byzantine Failure (MBF) model that can be plugged into the Analysis component to estimate how long the given application can be considered safe, and to trigger a global recovery procedure (through the Plan and Execute components) accordingly.

To this end, we model the temporal evolution of system compromise versus its recovery capabilities using \emph{Markov chains}.
We identify three models for such behavior, and analyze them using a \emph{Discrete Time Markov Chain} (DTMC) and \emph{Continuous Time Markov Chain} (CTMC), depending on their complexity.
In the CTMC we assume a known bound on the maximal attack rate $q$ on the system~\cite{SNVS06}, while the recovery rate $p$ is malleable by the system administrator.
In case the actual attack rate is smaller than $q$, the safety times computed in this paper are a lower bound for the true ones.

Leveraging the new proposed model, we present simple strategies for self-protection by predicting when should the entire system be recovered, and when can we still let local recoveries do the trick. 
Finally, we visualize the analytical results of the proposed model through a set of simulations.
The simulations code would be made available in open source once the anonymity requirement is~lifted.

The rest of the paper is organized as follows.
Section~\ref{sec:related} puts our approach into perspective with recent research attempts for related problems. Section~\ref{sec:model} gives a formal description of the execution model we consider and the probabilistic attacker model we introduce. Section~\ref{sec:architecture} describes the overall architecture of our proposal, 
\ifdefined\CONFversion
while Section~\ref{sec:analyzer} details the formal mathematical analysis part of the work, which forms a core component of our solution.
\else
while Sections~\ref{sec:analyzer}, \ref{sec:planner}, and \ref{sec:deployer} give further details about the key components at the core of our approach. 
\fi
Section~\ref{sec:simulations} presents simulations results to evaluate and visualize the feasibility of our scheme. Finally, Section~\ref{sec:conclusions} provides some concluding remarks.

\section{Related Work}
\label{sec:related}
\noindent{\bf Self-Protection Systems.} 
In the literature, various works have designed self-protecting systems to achieve different objectives~\cite{PEKARIC2023111716,10.1145/2555611,zhang2018self,Popov17,RS10,Lazarus,PBFT,LS04,SBNCNV10,SNVS06}. These systems often target specific domains. For instance, Yuan et al.~\cite{10.1145/2465478.2465479} and English et al.~\cite{english2006towards} propose self-protecting architectures: the former focuses on automatically detecting and mitigating software cyber threats, while the latter provides mitigation actions based on event correlation. Both approaches adapt the MAPE-K principles to enable dynamic runtime responses to security~threats.

Similarly, Liang et al.~\cite{8326530} introduce a framework to enhance the self-protection of power systems by leveraging the distributed security features of blockchain. In addition, other work focuses on securing data storage. 
For example, Strunk et al.~\cite{strunk2000self} aim to reduce the performance costs of system versioning, and Kocher et al.~\cite{kocher2003self} propose distributed watermarking algorithms to adaptively protect against piracy.
Preventing the spread of viruses across the different components of a distributed system has also been studied~\cite{nguyen14SRDSw,bonnet17sirocco}.
In particular, Nguyen et al.~\cite{nguyen14SRDSw} consider viruses that propagate and are detected probabilistically (countermeasures range from killing the detected compromised nodes or cutting the communication links leading to them). Then, Bonnet et al.~\cite{bonnet17sirocco} refined the concept by considering the communication topology and the relative speed of the countermeasure messages compared to the speed of the infection.

Sousa et al.~\cite{SNVS06} presented a model for analyzing the required elapsed time to rejuvenate the system assuming a maximal constant rate of failures and a maximal constant time that it takes to recover the entire system.
In their model, rejuvenation always involves the entire system rather than individual nodes as in our work.
Also, unlike Sousa et al.~\cite{SNVS06}, we assume a probabilistic failure propagation process rather than a constant failure rate.
Hence, the recovery time in Sousa et al.~\cite{SNVS06} is based on worst case timing assumption, while we aim for probabilistic guarantees.
The model of Sousa et al.~\cite{SNVS06} was extended~\cite{SBNCNV10} to combine both proactive and reactive rejuvenation methods to better ensure that the system remains correct at all times.

Several works (e.g.,~\cite{RS10,Lazarus,LS04}) studied how to ensure failure independence, e.g., by diversity.
In this work, we assume that either failures are independent, or that the infection probability per time step (in the discrete time Markov model) or average rate of failures (in the continuous time Markov model) are taken to represent the worst case scenario.
Ensuring such failure independence in beyond the scope of this work, i.e., we assume that some of the techniques presented in aforementioned work are used.

\smallskip

\noindent{\bf Mobile Byzantine Failure Models.}
Mobile Byzantine Failure (MBF) models have been introduced to capture a wide range of faults, including external attacks, virus infections, and arbitrary behavior caused by software bugs, within a unified framework that integrates detection and rejuvenation capabilities. In these models, failures are represented by an omniscient adversary capable of controlling up to $f$ mobile Byzantine agents. 
Each agent is stationed in a process, rendering it Byzantine faulty, until the adversary decides to move the agent to another process.
The key differences among existing MBF models lie in the power granted to the omniscient adversary (i.e., the conditions under which it can relocate agents) and the level of awareness each process has regarding its own failure state.

Most MBF models focus on \textit{round-based computations} and can be classified based on Byzantine mobility constraints. Under \emph{constrained mobility} \cite{DBLP:conf/ftcs/BuhrmanGH95}, the adversary can only move agents when protocol messages are exchanged, resembling the propagation of viruses. 
In contrast, under \emph{unconstrained mobility} \cite{banu2012improved,DBLP:journals/tcs/BonnetDNP16,DBLP:conf/wdag/Garay94,DBLP:conf/podc/OstrovskyY91,DBLP:conf/opodis/SasakiYKY13,DBLP:journals/iandc/Reischuk85}, agents do not move with messages, but rather at specific times.

In greater detail, Reischuk \cite{DBLP:journals/iandc/Reischuk85} examined scenarios in which malicious agents remain stationary for a certain period. 
Ostrovsky and Yung \cite{DBLP:conf/podc/OstrovskyY91} introduced the concept of mobile viruses and defined the adversary as an entity capable of injecting and distributing faults. 
Furthermore, Garay \cite{DBLP:conf/wdag/Garay94}, Banu et al. \cite{banu2012improved}, Sasaki et al. \cite{DBLP:conf/opodis/SasakiYKY13}, and Bonnet et al. \cite{DBLP:journals/tcs/BonnetDNP16} proposed models where processes perform computations in synchronous rounds and mobile agents can move between processes during a specific phase of the round. 
This mobility affects the ability of each process to adhere to the algorithm.
Consequently, the number of Byzantine faulty processes at any given time is limited; however, the identities of these Byzantine processes may change between consecutive rounds, and the effects of past compromises may persist if not adequately addressed.

The aforementioned works \cite{DBLP:conf/wdag/Garay94,banu2012improved,DBLP:conf/opodis/SasakiYKY13,DBLP:journals/tcs/BonnetDNP16} also differ in their assumptions about the knowledge processes possess about their prior infections.
In the Garay model \cite{DBLP:conf/wdag/Garay94}, a process can only detect its infection after the agent has left. 
In contrast, Sasaki et al. \cite{DBLP:conf/opodis/SasakiYKY13} explored a model in which processes are unable to detect when agents leave. 
Lastly, Bonnet et al. \cite{DBLP:journals/tcs/BonnetDNP16} proposed a middle ground, where non-faulty processes control the messages they send; specifically, they transmit the same message to all destinations and refrain from sending misleading information.

\section{System Model and Problem Statement}
\label{sec:model}
We consider a distributed system $\mathcal{S}$ consisting of $n$ distinct computing units, namely \textit{processes}. Each process $p_j$ is associated with a unique integer identifier $j$. All processes are able to interact directly with each other by exchanging messages on top of authenticated and reliable point-to-point links, as can be obtained, e.g., by running TLS over TCP/IP or QUIC.
In other words, nodes are connected by a communication network that can be abstracted by a complete graph where messages between two correct nodes are neither lost nor compromised while in transit over the links.

We consider the system to be synchronous (i.e., there exist upper bounds on the protocol execution time, on the time needed to exchange messages, and on the drift between physical clocks\footnote{For the ease of explanation we will assume that computation times and clock drift are negligible with respect to communication times and are assumed to be $0$.}). Under this assumption, we assume the existence of a fictional global clock that measures the passage of time.

Processes in the system cooperate to execute a distributed protocol $\mathcal{P}$ that implements a given specification describing their expected behavior. 
Each process $p_j$ has access to an external tamper-proof memory where it can store the genuine version of the distributed protocol $\mathcal{P}$. Processes may also have access to a small Trusted Execution Environment (TEE) that cannot be compromised by an external adversary, where they can run tiny handlers to manage specific messages.

We assume that processes may suffer \emph{Mobile Byzantine Failures}, i.e., while executing the protocol $\mathcal{P}$, processes could be compromised by an external adversary and may start behaving arbitrarily (i.e., they are prone to Byzantine failures), but the adversary is not able to compromise the code running on the TEE. 
Processes are also equipped with detection capabilities and, if they succeed in detecting their own misbehavior, they recover by retrieving the genuine code of $\mathcal{P}$ from their tamper-proof memory, and a correct state from other processes.

The execution of the protocol $\mathcal{P}$ passes through a set of \emph{system configurations} $\mathit{conf_0}, \mathit{conf_1}, \dots \mathit{conf_k} \dots$.
A system configuration $\mathit{conf_k}$ (\emph{configuration} for short) is defined as the set of pairs $\langle p_j, sec_j \rangle$ where $p_j$ is a process that is currently in the distributed system (and is currently executing the protocol $\mathcal{P}$) and $sec_j$ is a set of security features (e.g., security vulnerabilities and defense mechanisms) associated~to~$p_j$.

In this paper, we assume that processes in the distributed system are not \emph{diverse}, i.e., they are all characterized by the same set of security features and do not implement any form of diversity in the deployment.


Due to the presence of the external adversary, at any time $t$ a process $p_i$ can be in one of the following states: \textit{correct}, 
or \textit{faulty}.
A process is said to be \textit{faulty} at time $t$ if it is compromised and controlled by the adversary, 
while it is said to be \textit{correct} otherwise.

We assume that at time $t_0$, when the protocol starts, all processes in the system are \textit{correct}.

We assume that $\mathcal{P}$ is tolerant to a finite number $f$ of simultaneous Byzantine behaviors, i.e., $\mathcal{P}$ is guaranteed to behave correctly if and only if at any time $t$ the number of correct processes is at least than $n-f$ and we call the maximum number of tolerated failures \emph{protocol resilience} (or simply \emph{resilience}).
\smallskip

\noindent{\bf Problem statement.}
Ideally, we would like to predict how long it may take the system to reach a bad configuration where the number of Byzantine processes exceeds $f$.
Such a prediction can be used, for example, to initiate a periodic full manual restart of the entire system.

Similarly, if the number of Byzantine processes has exceeded $f$, we may want to predict if and how long it would take the system to return to a good configuration where the number of Byzantine nodes has dropped below $f$.
This prediction, in contrast, can be used to decide whether it makes sense to let the system recover on its own, or whether we need to manually restart it.
In particular, whenever $\mathcal{P}$ realizes a self-stabilizing Byzantine tolerant protocol~\cite{DD05,DRS23}, and the system is predicted to return quickly to a good configuration, not having to restart the entire system is highly beneficial.


\section{A Self-Recovering Architecture}
\label{sec:architecture}
In this section, we discuss an architecture implementing the MAPE-K paradigm that allows an existing distributed system running a protocol $\mathcal{P}$ to be automatically reconfigured to match its resilience threshold.

\begin{figure}[htbp]
    \centering
    \includegraphics[width=0.95\linewidth]{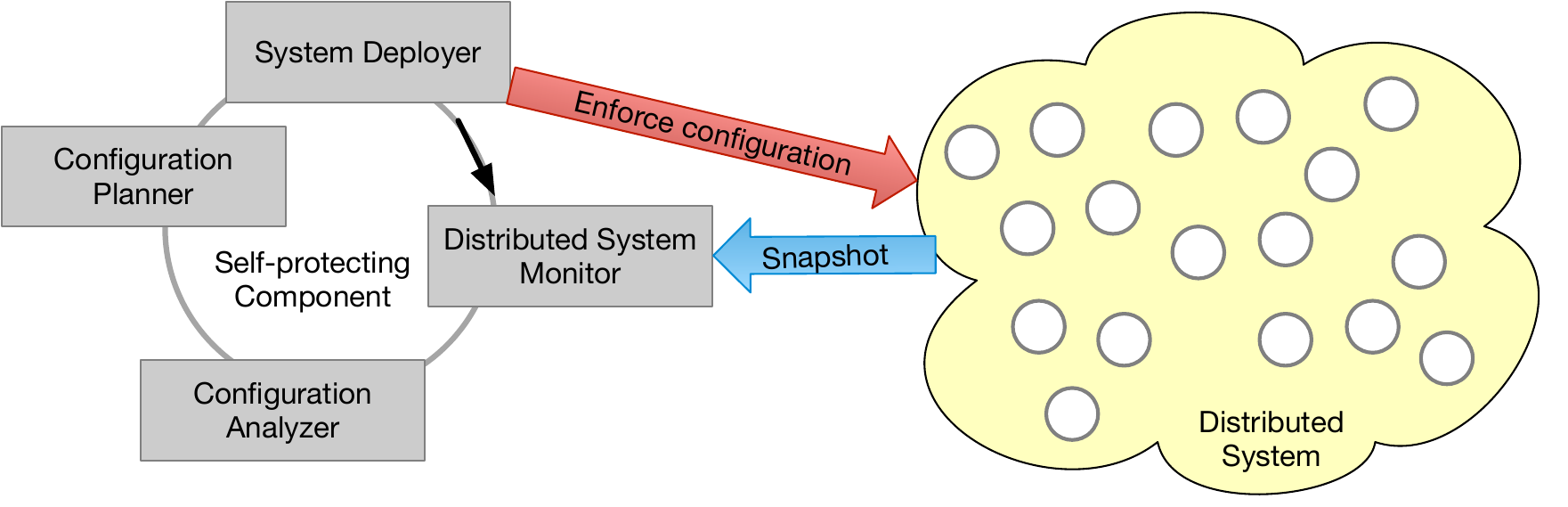}
    \caption{An Architecture for Self-protecting distributed systems with automatic reconfiguration.}
    \label{fig:architecture}
\end{figure}

Our architecture is depicted in Figure \ref{fig:architecture}, and it is composed of four main components:
\begin{LaTeXdescription}
    \item[Distributed System Monitor:] It is a module that continuously monitors the distributed system $\mathcal{S}$ and collects relevant information from it. 
    In particular, we are interested in constructing a \emph{snapshot} of the underlying system that reports the set of active processes and information about their current \textit{security settings} (e.g., the set of running applications and vulnerabilities affecting the process). 
    \item[Configuration Analyzer:] It is a module designed to estimate the security state of the current snapshot. 
    In particular, this module is responsible for estimating how long the system may still behave correctly with $\mathcal{P}$ running on enough correct processes.
    \item[Configuration Planner:] It is a module that has the goal of estimating if and when a system reconfiguration should occur to globally restore all the processes and bring $\mathcal{S}$ back to a configuration where all processes are correct.
    \item[System Deployer:] It is the component responsible for enforcing reconfiguration on the distributed system.
\end{LaTeXdescription}

Concerning the Distributed System Monitor component, several state-of-the-art techniques can be implemented by deploying network and vulnerability scanners to reconstruct and report a snapshot of the current distributed system configuration.
\ifdefined\CONFversion
The Configuration Planner's functionality is fairly simple; it is called by the Configuration Analyzer each time there is a new estimate for how long the system will remain in a safe state.
The Configuration Planner sets a corresponding timer, which if expires, invokes the establishment of a new configuration.
\fi
Similarly, the System Deployer can be implemented by encapsulating the protocol $\mathcal{P}$ into software containers that can be stored in the tamper-proof memory, run on a TEE at each process, and dynamically re-deployed in the distributed system environment.
\ifdefined\CONFversion
Thus, in the following, we focus primarily on the Configuration Analyzer.
In particular, we present a mobile Byzantine model, and analyze it through Markov chains.
Our analysis serves as the basic for predicting how long can we let the system run on its own, and try to fix problems using relatively cheap local rejuvenations, and when it is better to conduct a coordinated full system recovery.
\else
Thus, in the following, we will focus primarily on the Configuration Analyzer and Configuration Planner components.
In particular, we present a mobile Byzantine model, and analyze it through Markov chains.
Our analysis serves as the basic for predicting how long can we let the system run on its own, and try to fix problems using relatively cheap local rejuvenations, and when it is better to conduct a coordinated full system recovery.
\fi

\section{Configuration Analyzer}
\label{sec:analyzer}
The \emph{Configuration Analyzer} (CA) module takes as input a \emph{snapshot} of the monitored distributed system $\mathcal{S}$ (i.e., the current \emph{configuration} $\mathit{conf}_k$) and the \emph{resilience threshold} $f$ (i.e., the number of compromised processes that the protocol $\mathcal{P}$ can tolerate). Then, the CA outputs an estimation of the time period $\Delta_\mathit{safe}$ during which the distributed system remains correct (i.e., for which the attacker is not able to compromise more than $f$ processes, with high probability).

We model the distributed system $\mathcal{S}$ as a Markov chain to build our CA.
To aid understanding, we begin with a simplified case by describing the system as a discrete-time Markov chain in Section~\ref{sec:dtmc} and then generalize by utilizing the more expressive continuous-time model in Section~\ref{sec:ctmc}.

\subsection{Discrete Time Markov Chain}
\label{sec:dtmc}

Here, we model the distributed system $\mathcal{S}$ as a \emph{Discrete Time Markov Chain} (DTMC), whose state at time $t$ is denoted by $S_t$, where each state of the Markov chain represents the number of faulty processes in the system.
That is, assuming a system of $n$ processes, the model consists of $n+1$ states $s_0, s_1 \dots s_n$ (for the ease of notation, we alternately refer to a state $s_i$ simply with the number $i$ of compromised processes it represents).

Transitions in the Markov chain represent the compromise (due to attacks) or recovery (due to misbehavior detection) of processes, and are defined as follows. At any time $t$:
\begin{itemize}
    \item We move from a state $i$ to a state ${i+1}$ with probability $q_i=\mathbb{P}(S_{t}=i+1|S_{t-1}=i)$ (i.e., $q_i$ represents the probability that a new process is compromised at time $t$).
    \item We move from a state $i$ to a state ${i-1}$ with probability $p_i=\mathbb{P}(S_{t}=i-1|S_{t-1}=i)$ (i.e., $p_i$ represents the probability that a process is restored at time $t$).
    \item We remain in state $i$ with probability $r_i$ (i.e., $r_i$ represents the probability that no change in the number of faulty processes occurs).
\end{itemize}

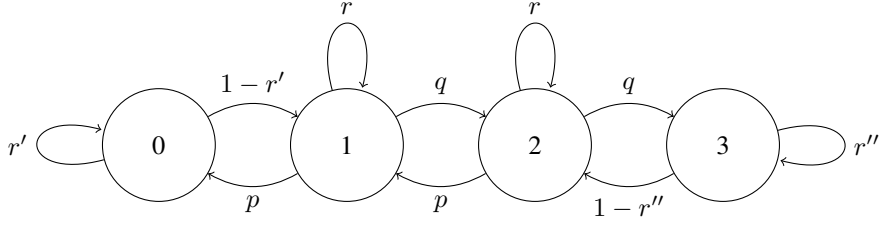
\begin{figure*}[t]
    \centering
\begin{tikzpicture}[
  node distance=2.5cm,
  state/.style={circle, draw, minimum size=1.5cm},
]

\node[state] (0) {0};
\node[state, right of=0] (1) {1};
\node[state, right of=1] (2) {2};
\node[state, right of=2] (3) {3};

\draw[->] (0) edge [loop left] node {$r'$} (0);
\draw[->] (1) edge [loop above] node {$r$} (1);
\draw[->] (2) edge [loop above] node {$r$} (2);
\draw[->] (3) edge [loop right] node {$r''$}(3);
\path[->] (0) edge [bend left] node [above] {$1-r'$} (1);
\path[->] (1) edge [bend left] node [below] {$p$} (0);
\path[->] (1) edge [bend left] node [above] {$q$} (2);
\path[->] (2) edge [bend left] node [below] {$p$} (1);
\path[->] (2) edge [bend left] node [above] {$q$} (3);
\path[->] (3) edge [bend left] node [below] {$1-r''$} (2);

\end{tikzpicture}
\caption{Example of a DTMC modelling a distributed system with 3 processes}
\label{fig:baseEx1}
\end{figure*}

For simplicity, here we assume that $\forall i>0$, $p_i = p$ and $\forall i<n$, $q_i = q$ for some configuration parameters $p$ and $q$.
In Subsection~\ref{sec:ctmc} below, we generalize this assumption as well.
Also, our analysis assumes either failure independence between process, e.g., by utilizing diversity techniques similar to the ones mentioned , e.g., in~\cite{RS10,Lazarus,LS04}, or by having $q$ represent the expected worst case scenario.

Figure~\ref{fig:baseEx1} shows an example of the DTMC model for a system of $3$ processes.
Leveraging this DTMC, we can estimate, for each snapshot provided by the distributed system monitoring component, the probability of being in a certain state $s_i$ at time $t$. Hence, we can estimate how long it would take to reach a state $s_{fail}$ where the number of correct processes falls below the resilience threshold for protocol $\mathcal{P}$.
The pseudo-code for the CA module is reported in Algorithm~\ref{alg:analyser}.

\begin{algorithm}
\footnotesize
\caption{Configuration Analyzer Algorithm\label{alg:analyser}}
\KwIn{Current system configuration ${\mathit{conf}_i}$}
\KwIn{Resilience threshold $f$ for protocol $\mathcal{P}$}
\KwOut{Estimated time $\Delta_\mathit{safe}$ to get to the state $s_{f+1}$}
\BlankLine
\BlankLine

{\bf upon event} ${\sf new\_snapshot}(\mathit{conf}_i, f)$\\
    $p \leftarrow {\sf estimate\_restore\_probability}(\mathit{conf}_i)$\;
    $q \leftarrow {\sf estimate\_infection\_probability}(\mathit{conf}_i)$\;
    $r \leftarrow {\sf estimate\_stability\_probability}(\mathit{conf}_i)$\;
    $\Delta_\mathit{safe} \leftarrow {\sf DTMC}(p, q, r, f)$\;
    {\bf trigger} ${\sf new\_estimation}(\Delta_\mathit{safe})$\;
        
 
\end{algorithm}

Let us note that, from a mathematical standpoint, the Markovian process proposed for the distributed system $\mathcal{S}$ forms a simple random walk.

Also, the probabilities $p$, $q$, and $r$ can be computed starting directly from the information included in the system configuration and are strictly dependent on the actual deployment.
For example, $q$ can be estimated based statistical averaging of observations over long term operation, while $p$ can be estimated by profiling detection tools like Intrusion Detection Systems deployed in the system.
How these features are actually implemented is beyond the scope of this paper.
Obviously, $p+q+r = 1$.
Having a value of $r>0$ enables us to accommodate for arbitrarily low values of $p$ and $q$.
From a theoretical standpoint, it simply adds a factor of \emph{laziness} to the random walk process, as discussed below.

In the edge case of $s_0$, all nodes are correct.
Hence, in this special case $p_0=0$, and we set $r_0=r'=r+p$.
Similarly, in the case of $s_n$, all nodes are Byzantine.
Hence, $q_n = 0$ and $r_n = r'' = r+q$.
When $n$ is large, the impact of these edge cases remains marginal.
\smallskip

\noindent {\bf DTMC Model Analysis.}
In the following, we study how the proposed DTMC model represents the evolution of the distributed system.
In particular, we are interested in how long it would take the system, starting from a given state (e.g., $s_0$ everyone is correct), to reach a state where a certain threshold of processes is faulty.
This is important, as this indicates how long we can let the system run on its own, without having to perform a manual hard reset.

Similarly, we want to investigate how long it would take for the system to cure itself once the threshold has been reached.
That is, if the system starts in a certain state where more than a threshold number of processes are faulty, we want to estimate how long the system should run until the number of faulty processes drops below the threshold again.
If this period is acceptable, we may decide to wait and let the system cure itself without manual intervention.
Otherwise, we would immediately hard reset the system.
%
%

For didactic purposes, we start with analyzing simple cases and then generalize the results.
Below, Theorem~\ref{thm:simple} handles the case in which infection and recovery have equal probabilities, Theorem~\ref{thm:high-recover} covers the case in which the recovery probability is higher, while Theorem~\ref{thm:high-infection} deals with higher infection probability.
In the following theorem, $c$ represents the threshold of faulty processes we are interested in; $c$ is a fraction between $0$ and $1$. For any state $x$ of DTMC let $\tau_x$ be the time to reach state $x$, that is
\ifdefined\CONFversion
$\tau_x = \mathrm{inf}\{t>0 : S_t=x\}$.
\else
$$\tau_x = \mathrm{inf}\{t>0 : S_t=x\}.$$
\fi
We denote $$\mathbb{E}_x[\cdot]=\mathbb{E}[\cdot|S_0=x].$$
\begin{theorem}
\label{thm:simple}
 If $q=1-p$, $p=\frac{1}{2}$, and $r=0$ then the expected time to reach state $cn$ is $\mathbb{E}_0[\tau_{cn}]=(cn)^2$.
\end{theorem}

\begin{proof}
This case corresponds to a simple random walk over the chain.
Hence, we have $$\tau_x = \mathrm{inf}\{t>0 : S_t=x\}.$$

The following process is a martingale~\cite{durrett2019probability}, $S^2_t - t$.
Our goal is to find $\mathbb{E}[\tau_{cn}]$.
By the optional stopping theorem, we have:
\ifdefined\CONFversion
$$\mathbb{E}_0[S^2_{\tau_{cn}} - \tau_{cn}]=\mathbb{E}[S^2_{0} - 0]=0. \quad \mathbb{E}_0[\tau_{cn}] = (cn)^2$$
\else
$$\mathbb{E}_0[S^2_{\tau_{cn}} - \tau_{cn}]=\mathbb{E}[S^2_{0} - 0]=0.$$
$$\mathbb{E}_0[\tau_{cn}] = (cn)^2$$
\fi
and the claim follows.
\end{proof}

\begin{theorem}
\label{thm:high-recover}
If $q=1-p$, $p > \frac{1}{2}$, and $r=0$ then the expected time to reach state $cn$ is bounded below by $\mathbb{E}_0[\tau_{cn}]\ge \frac{1}{2}\left(\frac{p}{q}\right)^{\frac{n}{3}}$.
\end{theorem}

\begin{proof}
A good estimate follows the gambler's ruin probability~\cite{bertsekas2008introduction}. $\mathbb{P}^1(\tau_{n/3}<\tau_0)=\frac{\frac{p}{q}-1}{\left(\frac{p}{q}\right)^{\frac{n}{3}}-1}<
    \frac{2}{\left(\frac{p}{q}\right)^{\frac{n}{3}}}$. 
    Thus, in order to reach $n/3$, the random walk needs to make at least Geometric$\left(\frac{2}{\left(\frac{p}{q}\right)^{\frac{n}{3}}}\right)$ attempts. 
    Each attempt requires at least one step. Thus, we obtain
\ifdefined\CONFversion
    $\mathbb{E}_0[\tau_{cn}]\ge \frac{1}{2}\left(\frac{p}{q}\right)^{\frac{n}{3}}$.
\else
    $$
    \mathbb{E}_0[\tau_{cn}]\ge \frac{1}{2}\left(\frac{p}{q}\right)^{\frac{n}{3}}
    .$$
\fi
\end{proof}

Theorem~\ref{thm:high-recover} shows that if recovery is faster than infection, reaching the failure threshold takes exponential time, suggesting that local rejuvenation is sufficient under these conditions.

\begin{theorem}
\label{thm:high-infection}
 If $q=1-p$, $p < \frac{1}{2}$, and $r=0$ then the expected time to reach state $cn$ is bounded below by  $\mathbb{E}_0[\tau_{cn}]\ge \frac{n}{1-2p}\left(1-\frac{p}{q}\right)^{-1}$.
\end{theorem}

\begin{proof}
In this case, by gamblers ruin, $\mathbb{P}^1(\tau_{n/3}<\tau_0)=\frac{\frac{p}{q}-1}{\left(\frac{p}{q}\right)^{\frac{n}{3}}-1}\approx
    1-\frac{p}{q}$. 
    Thus, the number of excursions from $0$ is bounded in expectation and the expected hitting time is
    $$
    \mathbb{E}_1[\tau_{cn}]\ge \mathbb{E}_1[\tau_{cn}|\tau_{cn}<\tau_0]\mathbb{P}^1(\tau_{cn}<\tau_0)^{-1}$$
    $$\ge\frac{n}{1-2p}\left(1-\frac{p}{q}\right)^{-1}
    .$$
\end{proof}

\smallskip
\noindent \textbf{Generalizing for $r>0$ ($\forall i, p_i=p, q_i=q$).}
In this case, the time is simply multiplied by $\frac{1}{1-r}$.
This is because now it takes on average $\frac{1}{1-r}$ times longer to make each state transition (in either direction), as the probability to make a state transition (rather than staying in the same place) is $1-r$.

\subsection{Continuous Time Markov Chain}
\label{sec:ctmc}

The \emph{Continuous Time Markov Chain} (CTMC) model is similar to the DTMC model.
Here too, each state of the Markov chain represents the number of faulty processes in the system (i.e., assuming a system of $n$ processes, the model consists of $n+1$ states $0, 1 \dots n$).
Further, transitions in the Markov chain represent the compromise (due to attacks) or recovery (due to misbehavior detection) of processes.
However, since time is continuous, transitions can occur at any arbitrary moment, so the parameters $p_i$ and $q_i$ on the edges represent the average rate in which they occur, taken from the exponential distribution (we present three models for computing $p_i$ and $q_i$~below):
\begin{itemize}
    \item We move from a state $i$ to a state ${i+1}$ with a rate of $q_i$. That is, $q_i$ represents the average rate in which a new process is compromised given that there are already $i$ compromised nodes.
    \item We move from a state $i$ to a state ${i-1}$ with a rate $p_i$. That is, $p_i$ represents the average rate by which a process is restored whenever there are $i$ compromised nodes.
\end{itemize}

We distinguish between three sub-models of assigning the rates $p_i$s and $q_i$s, representing three attackers and recovery modes, all of which are instantiated by a pair of configuration parameters $p$ and $q$, as described below:

\begin{LaTeXdescription}
    \item[External:] In the external model, for all values of $i>0$, $p_i = p$ and for all $i<n$, $q_i = q$.
    This model is analogous to the one analyzed in Subsection~\ref{sec:dtmc} above.
    We remind the reader that this model is mostly presented here for didactic purposes.
    
    \item[Internal:] In the internal model, attacks are initiated independently by the Byzantine nodes themselves. Each Byzantine node attacks another node at a rate $q$. However, it might try to attack a node that is already Byzantine, which is pointless.
    Thus, only $n-i$ nodes out of $n$ contribute to the infection rate. Hence,  $q_i = q \times i \times \frac{n - i}{n}$.

    Nodes recover themselves in an independent manner as well at an average rate of $p$.
    Yet, obviously, if a healthy node attempts to recover itself, it does not change the number of non-Byzantine nodes.
    Hence, $p_i = p \times i$.
    
    \item[Coordinated:] In the coordinated mode, attacks by initiated by the Byzantine nodes as well, but this time in a coordinated fashion, meaning that they avoid trying to attack an already infected node.
    Hence, here, $q_i = q \times i$.

    Healthy nodes behave as in the Internal model.
    That is, $p_i = p \times i$.
\end{LaTeXdescription}

We note that in the Internal and Coordinated models, it is possible that $p_i + q_i > 1$.
This is OK since in CTMC, $p_i$ and $q_i$ represent rates rather than probabilities.
Hence, whenever $p_i + q_i > 1$, it simply means that on average there can more more than one state transition per one time unit.
Also, our analysis assumes either failure independence between process, e.g., by utilizing diversity techniques similar to the ones mentioned, e.g., in~\cite{RS10,Lazarus,LS04}, or by having $q$ represent the expected worst case intrusion rate~\cite{SNVS06}.
We emphasize that the analysis is correct for any values of $p$ and $q$, and in particular, the case where $p=0$, in which the system is defenseless.

We note that many realistic systems, including, e.g., computer networks, are often modeled by continuous time poison processes of fixed rates.
The standard explanation is that as long as asymptotically the number of packet arrivals in networking, or attacks in our case, is independent between disjoint time intervals, the asymptotic average rate adequately describe the entire model~\cite[Chapter 5]{bertsekas2008introduction}.
The Poisson process super-positioning property also takes care of variations between individual machines.

\subsubsection{CTMC-External Model Analysis}

Notice that the CTMC-External model is mathematically equivalent to the model that we analyzed under DTCM (Subsection~\ref{sec:dtmc}).
Hence, the results apply the same when assigning each time step to a single time unit.

\subsubsection{CTMC-Internal Model Analysis}
\label{sec:CTMC-Internal-Analysis}
\paragraph{Time to reach first bad configuration}
Here, moving left (rejuvenation without infection) occurs with probability $\frac{pi}{pi+qi(n-i)/n}$ while moving right (infection without rejuvenation) occurs with probability $\frac{qi(n-i)/n}{{pi+qi(n-i)/n}}$.
I.e., a step in either direction occurs every $\min\{\text{exp}(px),\text{exp}(qx(n-x)/n)\}$~time.

\begin{theorem}
    There exists $c$ such that the expected time to reach the threshold $n/3$ faulty nodes, assuming initially all nodes are correct, can be estimated as:
    $$\mathbb{E}_0[\tau_{\frac{1}{3}n}]\approx \left\{\begin{array}{ll}
         \frac{1}{p\frac{n}{3}+q\frac{5n}{18}} e^{cn} & p > \left(\frac{2}{3}+\epsilon\right)q \\
         c\log(n) & p < \frac{2}{3}q
    \end{array}\right.$$
\end{theorem}

\begin{proof}
We define $f(i)= \mathbb{E}_i[\tau_{cn}]$, with $f(cn) = 0$, $f(0)=f(1)+1$.
We have 
\begin{equation}\begin{aligned}f(i) &= \frac{qi(n-i)/n}{pi+qi(n-i)/n}f(i+1) + \frac{pi}{pi+qi(n-i)/n}f(i-1) \\&+ \frac{1}{pi+qi(n-i)/n}.\end{aligned}\end{equation}

\paragraph*{Private case of $p=0$} Here, we have
$\Sigma_{x=1}^{cn} \frac{1}{qx}\frac{n}{n-x} = \frac{1}{q}\log(cn)$, which matches known results from epidemiology and probabilistic broadcast theory.

\paragraph*{Large $p$} Whenever $p>q\frac{n-i}{n}$ then there is a drift to the left (recovery).
Since $\frac{n-i}{n}$ is monotonically decreasing in $i$, the drift to the right in the interval $\left[\frac{1}{6}n,\frac{1}{3}n\right]$ is uniformly bounded above by $\frac{5}{6}q$. 
Thus, under the assumption $p>\frac{5}{6}q$ in the interval $\left[\frac{1}{6}n,\frac{1}{3}n\right]$ the process has total drift to the left. 
Moreover, inside this interval, the expected time of each step in the interval, is at least $\frac{1}{p\frac{n}{3}+q\frac{n}{3}\frac{n-n/6}{n}}=\frac{1}{p\frac{n}{3}+q\frac{5n}{18}}$. 
The conclusion is that the expected time that the Markovian process goes from $\frac{1}{6}n$ to $\frac{1}{3}n$ is at least $\frac{1}{p\frac{n}{3}+q\frac{5n}{18}} e^{cn}$. 

Note that we could have taken $p>\left(\frac{2}{3}+\epsilon\right)q$, and the Markovian process would have a drift to the left in a small linear interval $[(\frac{1}{3}-\epsilon)n,\frac{1}{3}n]$. Thus, one would still get asymptomatically an exponential time to reach the threshold, though with a smaller constant in the exponent.

\paragraph*{Small $p$} Under the assumption $p<\frac{2}{3}q$ for every $i\le \frac{1}{3}n$, there is a uniformly positive drift to the right. 
The Markovian process reaches $\frac{1}{3}n$ in an expected $cn$ steps. The expected time to make these steps is: 
\begin{equation}
\mathbb{E}_0[\tau_{\frac{1}{3}n}]\approx \sum_{i=1}^{\frac{1}{3}n}\frac{c(p,q)}{pi+qi(n-i)/n}=c\log(n).
\end{equation}
\end{proof}

\paragraph{Distribution of time spent in each state}\label{sec:state-dist-external}
We show the proportion of time that the chain visits each state by analysis of the stationary distribution, denoted $\pi(i)$, of the Markov chain. The detailed balance equation, characterizing $\pi(i)$, is $\pi(i)q_i=\pi(i+1)p_i$. By arbitrarily choosing $\pi(1)=1$ (remembering to normalize later to get a probability distribution), we get for all $2\ge k\le n$
\small
$$\pi(k)=\left(\frac{q}{np}\right)^{k-1}\prod_{i=n-k}^{n-2}i.$$ 
\normalsize
In the case where $p>q$, $\pi(k)$ decays exponentially, thus the Markov chain will spend most of its time in the low infection regime. If $q>p$, then by calculating the critical point of $\pi(k)$, one gets the maximal occupation density at the $k$ satisfying $p=\frac{n-k}{n}q$. That is the location in which the infection and recovery rates coincide.

\subsubsection{CTMC-Coordinated Model Analysis}

In this model, we get for $f(i)= \mathbb{E}_i[\tau_{cn}]$, with $f(cn) = 0$ and~$f(0)=f(1)+1$,
$$f(i) = \frac{qi}{pi+qi}f(i+1) + \frac{pi}{pi+qi}f(i-1) + \frac{1}{pi+qi}.$$
Thus,
$$
f(i)-f(i+1)=\frac{p}{q}\left[f(i-1)-f(i)\right]+\frac{1}{qi}.
$$
Using the boundary conditions and denoting $\zeta=\frac{p}{q}$:
\small
\begin{equation}
\begin{aligned}
  &   f(0)-f(1)=1\\
   & f(1)-f(2)=\zeta( f(0)-f(1))+\frac{1}{q}=\zeta+\frac{1}{q}\\
 &  f(2)-f(3)=\zeta( f(1)-f(2))+\frac{1}{2q}=\zeta^2+\frac{\zeta}{q}+\frac{1}{2q}\\
  &  f(3)-f(4)=\zeta( f(2)-f(3))+\frac{1}{3q}=\zeta^3+\frac{\zeta^2}{q}+\frac{\zeta}{2q}+\frac{1}{3q}\\
\end{aligned}
\end{equation}
\normalsize
Now taking the telescopic sum up to $\frac{n}{3}$ and rearranging the order of sums, we obtain
\small
\begin{equation}
\begin{aligned}
f(1)=\sum_{k=1}^{\frac{n}{3}}\sum_{j=0}^{\frac{n}{3}-k-1}\frac{\zeta^j}{kq}-1=\\
\frac{c}{q(1-\zeta)}\log(\frac{n}{3})-\frac{c}{q(1-\zeta)}\sum_{k=1}^{\frac{n}{3}}\frac{1}{k}\zeta^{\frac{n}{3}-k}
\end{aligned}
\end{equation}
\normalsize
Hence, if $p<q$ and thus $\zeta<1$, we obtain an order of $\log(\frac{n}{3})$.
If $p>q$, and thus $\zeta>1$, we obtain an order of $\zeta^{\frac{n}{3}}$.

\section{Configuration Planner}
\label{sec:planner}
The \emph{Configuration Planner} (CP) module'goal is to estimate when a new configuration needs to be deployed and how such a configuration should be composed.
The CP module is triggered by the CA module and takes as input the estimate of the safety period $\Delta_\mathit{safe}$. It then computes when a global reconfiguration is needed, to rejuvenate all processes in $\mathcal{S}$, and bring the distributed system back to a state where no process is compromised. 
In this paper, we propose a simple implementation for the CP module (see Algorithm \ref{alg:step1}) that, given the time $\Delta_\mathit{reconfig}$ needed to effectively start a reconfiguration, computes \emph{(i)} when the next global reconfiguration should take place, and \emph{(ii)} when the System Deployer module should be activated.

\RestyleAlgo{ruled}

\begin{algorithm}
\footnotesize
\caption{Configuration Planner Algorithm\label{alg:step1}}
\KwIn{The current estimation of the safe time to reach a given state $s_x$ where faulty processes may exceed the resilience threshold $f$}
\KwOut{A new configuration $\mathit{conf}_j$ to be deployed}

\BlankLine
\BlankLine

{\bf Init}\\
$\mathit{timer} \leftarrow \infty$\;
 
\BlankLine
\BlankLine

{\bf upon event} ${\sf new\_estimation}(\Delta_\mathit{safe})$\\
\eIf{$\delta < \Delta_\mathit{safe}$}{
    $\mathit{timer} \leftarrow \Delta_\mathit{safe}-\delta$\;
}
{
    $timer \leftarrow 0$\;
}
\BlankLine
\BlankLine

{\bf upon event} ${\sf Timeout}()$\\
$\mathit{conf}_j \leftarrow {\sf define\_new\_configuration}()$\;
{\bf trigger}$ ~{\sf new\_configuration}(\mathit{conf}\_j)$\;

        
 
\end{algorithm}

The proposed algorithm works by leveraging the synchronous setting hypothesis, and uses timers to schedule reconfigurations, i.e., to generate a new configuration $\mathit{conf}_j$ that must be deployed. 
In the current algorithm, the new configuration $\mathit{conf}_j$ (computed by the ${\sf define\_new\_configuration}()$ function) is simply defined as the initial configuration $conf_0$, where all processes have been sanitized by retrieving the correct code from their tamper-proof memories.

\ifdefined\CONFversion
\else
\section{System Deployer}
\label{sec:deployer}

The \emph{System Deployer} (SD) module is responsible for enforcing a given configuration $\mathit{conf}_j$. In particular, it is in charge of actuating the reconfiguration strategy over the distributed system $\mathcal{S}$.

\begin{algorithm}
\footnotesize
\caption{System Deployer\label{alg:deployer}}
\KwIn{The new Configuration to be Deployed}

\BlankLine
\BlankLine

{\bf Init}\\
$\mathit{last}\_\mathit{configuration} \leftarrow \mathit{conf}_0$\;

\BlankLine
\BlankLine

{\bf upon event}$~{\sf new\_configuration}(\mathit{conf}_j)$\\
\ForEach{$p_i \in \mathit{last}\_\mathit{configuration}$}{
    {\bf trigger} $\sf {pp2p-send}(REBOOT, \mathit{conf}_j)$ to $p_i$\;
    }
$\mathit{last}\_\mathit{configuration} \leftarrow \mathit{conf}_j$\;

        
 
\end{algorithm}

When triggered by the CP module, the SD takes care of interacting with the Trusted Execution Environment running on process $p_i$ and communicates that a reboot is required to move to a new configuration $\mathit{conf}_j$ (see Algorithm~\ref{alg:deployer}).
\fi

\section{Simulations}
\label{sec:simulations}
\subsection{DTMC Simulations}
\label{sec:dtmc-simulations}

We now present simulations of the DTMC model.
The simulations extend the example given in Figure~\ref{fig:baseEx1} to any number of processes $n$, where $p$ is the probability that a single node is healed in a given step, $q$ is the probability that a single node is compromised, and $r$ is the probability of remaining in the same state (the number of infections is equal to the number of compromises).
Additional parameters include the total number of requested simulation steps, the initial state, and the threshold of Byzantine nodes below which the system is considered good and above which the system is considered bad.
For example, it is known that asynchronous and partially synchronous Byzantine consensus, as well as synchronous Byzantine consensus with oral messages, require a maximum of $f<n/3$ Byzantine nodes.
To mimic that, we can specify in the simulation that the threshold is just under~$1/3$.

The metrics we consider are \textit{(i)} the percentage of runs in which the system has remained in a good state, \textit{(ii)} the percentage in which the system was in a bad state, and \textit{(iii)} how long it took to make the first switch between good and bad, i.e., reach the threshold.
Each data point shown is an average of 100 runs.

Figure~\ref{fig:Heatmap-Good-200-66} shows the simulation results for ($i$) the percentage of runs in which the system remained in a good state (the number of Byzantine nodes remained below the threshold) during the entire run, ($ii$) the percentage of runs in which the system flipped from a good state to a bad state, and ($iii$) the time it took to switch for the first time from a good state to a bad state.
The presented results are for simulations taking 1M simulation steps with 200 nodes, a threshold of $1/3$ (66 nodes), in which the system started with 0 Byzantine nodes.

As expected from the theoretical analysis, whenever $p>q$, the time to reach the first flip is very large, taking beyond 1M steps when $n=200$.
Hence, whenever $p>q$, we get that all runs were purely good.
Let us note that by the definition of our DTMC model, the cells that represent $p+q>1$ are meaningless.
Finally, as expected from theory, whenever $p=q$ we always have a flip, and the time for the first flip is proportional to the value of $p$ (and $q$).

\begin{figure*}
    \centering
    \includegraphics[width=\linewidth]{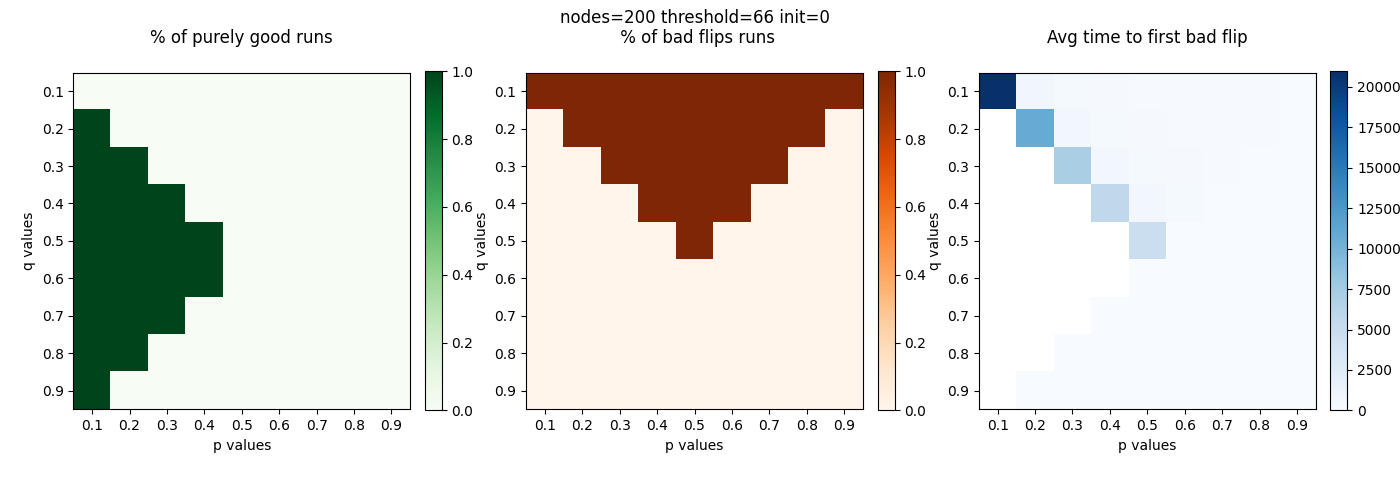}
    \caption{Percentage of totally good runs, percentage of runs in which the system flips from good state to bad state, and the average time taken for the system to flip to the first bad state in those runs, when the number of nodes is 200, the initial state is 0 (no Byzantine), the threshold is 1/3 (66 nodes), and the simulation is run for 1M steps.}
    \label{fig:Heatmap-Good-200-66}
\end{figure*}

Similarly, Figure~\ref{fig:Heatmap-Bad-200-66} shows the simulation results for ($i$) the percentage of runs in which the system remained in a bad state (the number of Byzantine nodes remained above the threshold) during the entire run, ($ii$) the percentage of runs in which the system flipped from a bad state to a good state, and ($iii$) the time it took to switch for the first time from a bad state to a good state.
This is for simulations taking 1M simulation steps with 200 nodes, threshold of $1/3$ (66 nodes), where the system started with 200 Byzantine nodes (all nodes are Byzantine).
As expected from the theoretical analysis, whenever $p<q$, the time to reach the first flip is very large, taking beyond 1M steps when $n=200$.
Hence, whenever $p<q$, we get that all runs were purely bad.
As before, by definition, the cells that represent $p+q>1$ are meaningless.
Finally, as expected from theory, whenever $p=q$ we always have a flip, and the time for the first flip is proportional to the value of $p$ (and $q$).
Here the times are longer compared to those shown in Figure~\ref{fig:Heatmap-Good-200-66} since when the threshold is $1/3$, there are about twice as many bad states as good states.

\begin{figure*}
    \centering
    \includegraphics[width=\linewidth]{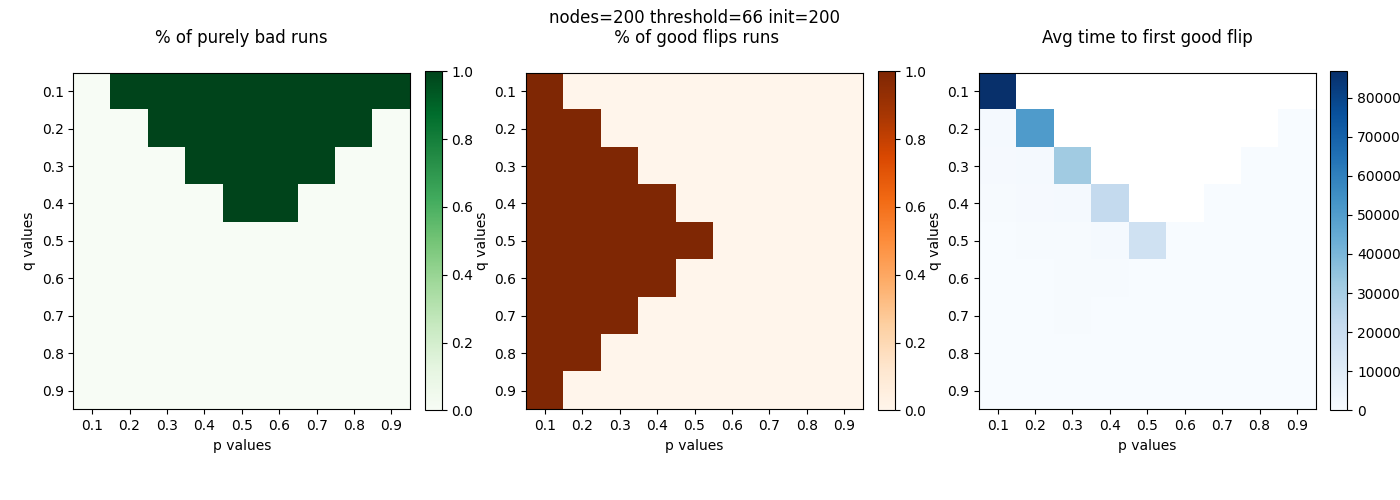}
    \caption{Percentage of totally bad runs, percentage of runs in which the system flips from bad state to good state, and the average time taken for the system to flip to the first good state in those runs, when the number of nodes is 200, the initial state is 200 (all Byzantine), the threshold is 1/3 (66 nodes), and the simulation is run for 1M steps.}
    \label{fig:Heatmap-Bad-200-66}
\end{figure*}

\subsection{CTMC Simulations}
\label{sec:ctmc-simulations}

We now present the simulation results for the three variants of the CTMC model, namely EXTERNAL, INTERNAL, and COORDINATED.
For \ifdefined\CONFversion lack of space, \else brevity, \fi 
we only show a subset of all results that are most interesting.
The important message is that the theoretical analysis is validated by the simulation results.
The full set of graphs will be made available in an accompanying manuscript published once the anonymity requirement is lifted.

Recall that in these models, $p$ and $q$ control the rate at which state transitions occur, and thus, whenever $p+q$ are large, it simply means that transitions happen faster than the equivalent of a single (discrete) time unit (by definition, time is continuous here).
Consequently, we now measure the time that it takes to cross the threshold, either from good to bad or vice versa, and the total runtime units the Markovian process remained in a given state $i$, or set of states, rather than the number of steps.
All parameters here are the same as before, except that the simulated running time was chosen to be 100,000, which usually translates into more than 100,000 due to the continuity of time here.

\subsubsection{EXTERNAL Results}
\label{sec:ctmc-external-simulations}

Unsurprisingly, the results here for the percentage of purely good runs, the percentage of bad flips runs, and the average time to the first bad flip when starting at state 0 are almost indistinguishable from what was shown for the DTMC model in Figure~\ref{fig:Heatmap-Good-200-66}.
Similarly, the results here for the percentage of purely bad runs, the percentage of good flips runs, and the average time to the first good flip when starting at state 200 are almost indistinguishable from what was shown for the DTMC model in Figure~\ref{fig:Heatmap-Bad-200-66}.
This is because the processes are mathematically analogous.

\begin{figure*}[t]
    \centering
    \begin{subfigure}{0.3\textwidth}
    \includegraphics[width=\textwidth]{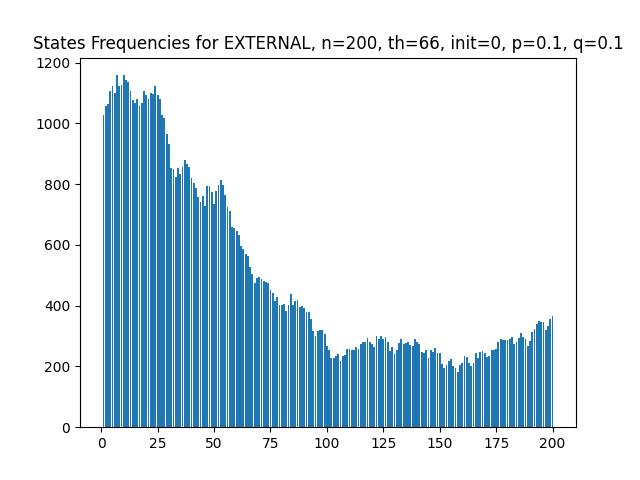}
    \caption{State distribution $p=q=0.1$}
    \label{fig:external-distribution:01-01-0}
    \end{subfigure}
    \begin{subfigure}{0.3\textwidth}
    \includegraphics[width=\textwidth]{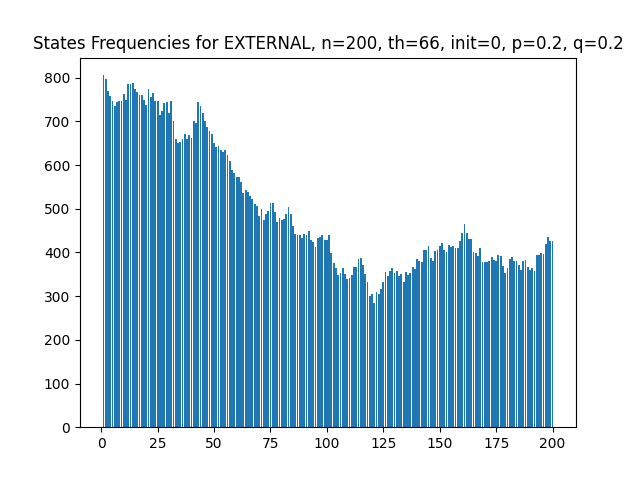}
    \caption{State distribution $p=q=0.2$}
    \label{fig:external-distribution:02-02-0}
    \end{subfigure}
    \begin{subfigure}{0.3\textwidth}
    \includegraphics[width=\textwidth]{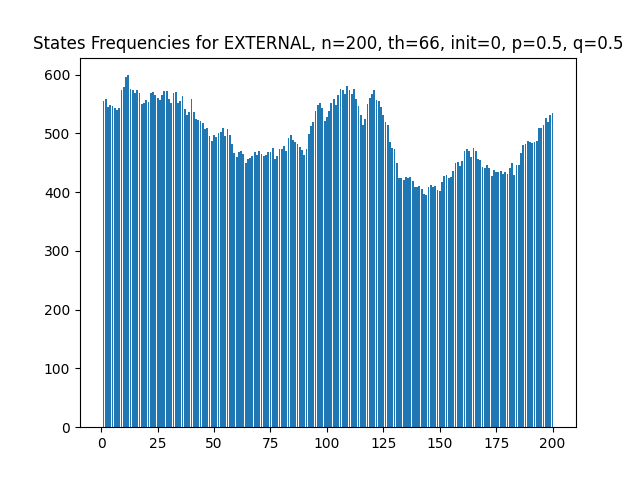}
    \caption{State distribution $p=q=0.5$}
    \label{fig:external-distribution:05-05-0}
    \end{subfigure}
    \begin{subfigure}{0.3\textwidth}
    \includegraphics[width=\textwidth]{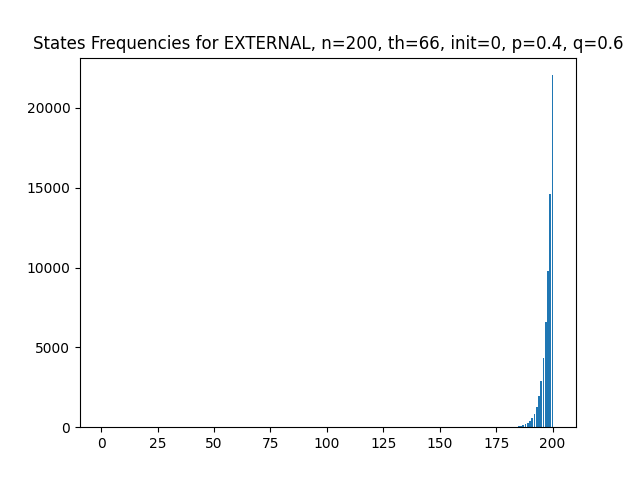}
    \caption{State distribution $p=0.4, q=0.6$}
    \label{fig:external-distribution:04-06-0}
    \end{subfigure}
    \begin{subfigure}{0.3\textwidth}
    \includegraphics[width=\textwidth]{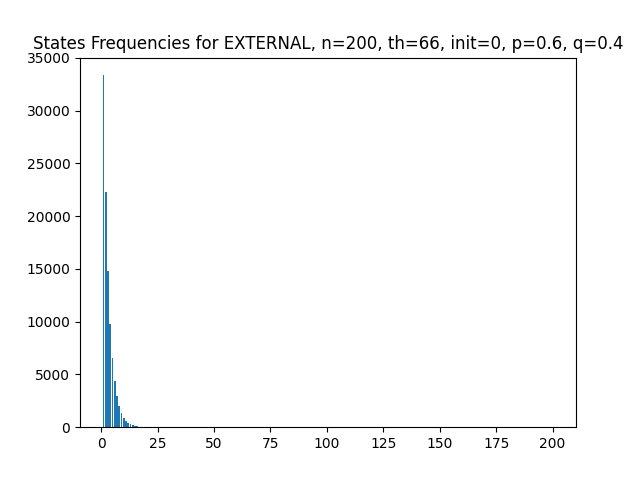}
    \caption{State distribution $p=0.6, q=0.4$}
    \label{fig:external-distribution:06-04-0}
    \end{subfigure}
    \caption{Distribution of time spent at different states when the number of nodes is 200, the initial state is 0 (no Byzantine), and the simulation is run for 100K steps for the EXTERNAL model.}
    \label{fig:external-distribution}
\end{figure*}

Figure~\ref{fig:external-distribution} illustrates the percentage of time spent in each state for given values for $p$ and $q$, and a given starting state $i$, with $200$ nodes.
As can be seen, when $p=q=0.5$ (Figure~\ref{fig:external-distribution:05-05-0}), the distribution of times is almost uniform.
On the other hand, when $p=q=0.2$ and we start at state $0$ for example (Figure~\ref{fig:external-distribution:02-02-0}), we see a slope that goes down initially until leveling.
The reason for this slope is that sine $p$ and $q$ are smaller, it takes more time for the Markovian process to make enough steps to reach its stationary distribution (which is uniform).
However, as soon as $p>q$, even by a small margin, as shown in Figure~\ref{fig:external-distribution:06-04-0}, most of the time the Markovian process remains at state $0$ or close to it.
Alternatively, as soon as $p<q$, even by a small margin, as shown in Figure~\ref{fig:external-distribution:04-06-0}, most of the time the Markovian process remains at state $200$ or close to it, even when starting at $0$.
These phenomena become more extreme as the difference between $p$ and $q$ grows.

\subsubsection{INTERNAL Results}
\label{sec:ctmc-internal-simulations}

Figure~\ref{fig:internal-distribution} illustrates the percentage of time spent in each state for given values for $p$ and $q$, and a given starting state $i$, with $200$ nodes.
Here, we can see an interesting phenomenon: even when $q>p$, the Markovian process spends most of its time close to some intermediate value, depending on the exact ratio between $p$ and $q$.
This is fairly independent of the initial state, as can be seen when comparing Figures~\ref{fig:internal-distribution:02-08-0} vs.~\ref{fig:internal-distribution:02-08-66} and Figures~\ref{fig:internal-distribution:03-07-0} vs.~\ref{fig:internal-distribution:03-07-66} (and the longer the run would be, the smaller the impact of the initial state would become).
Once more, this corresponds well to the mathematical analysis in Section~\ref{sec:CTMC-Internal-Analysis}.
For example, consider the case of $p=0.4, q==0.6$ (Figure~\ref{fig:internal-distribution:04-06-0}).
The analysis in Section~\ref{sec:CTMC-Internal-Analysis} calculates that the maximal occupation density would be at the $i$ satisfying $p=\frac{n-i}{n}q$.
In other words, the corresponding $i$ is $n(1-\frac{p}{q})$.
In this specific example, we get $i=200\times(1-\frac{0.4}{0.6}) = 66$.
In other words, in this case, even though the individual rate of infection is larger than the individual rate of self-recovery, the system stabilizes to a state in which most nodes are correct!

\begin{figure*}[t]
    \centering
    \begin{subfigure}{0.3\textwidth}
    \includegraphics[width=\textwidth]{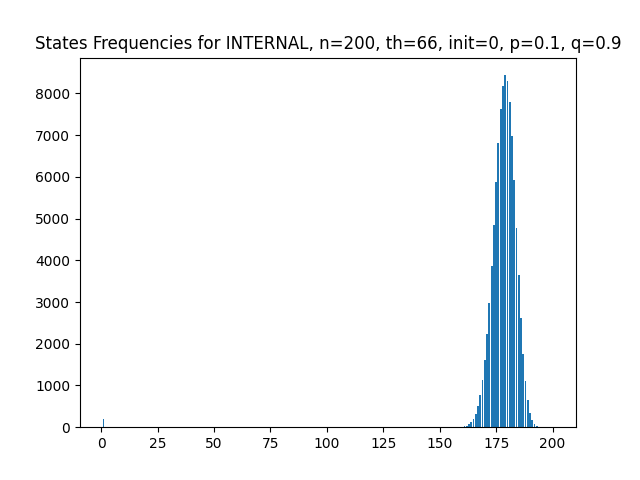}
    \caption{State distribution $p=0.1, q=0.9$}
    \label{fig:internal-distribution:01-01-0}
    \end{subfigure}
    \begin{subfigure}{0.3\textwidth}
    \includegraphics[width=\textwidth]{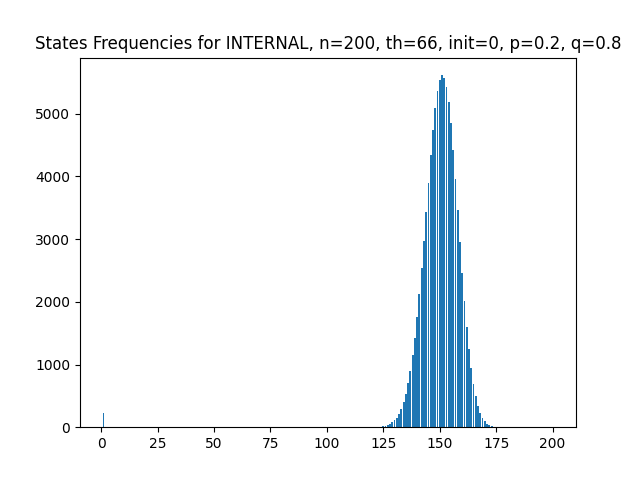}
    \caption{State distribution $p=0.2, q=0.8$}
    \label{fig:internal-distribution:02-08-0}
    \end{subfigure}
    \begin{subfigure}{0.3\textwidth}
    \includegraphics[width=\textwidth]{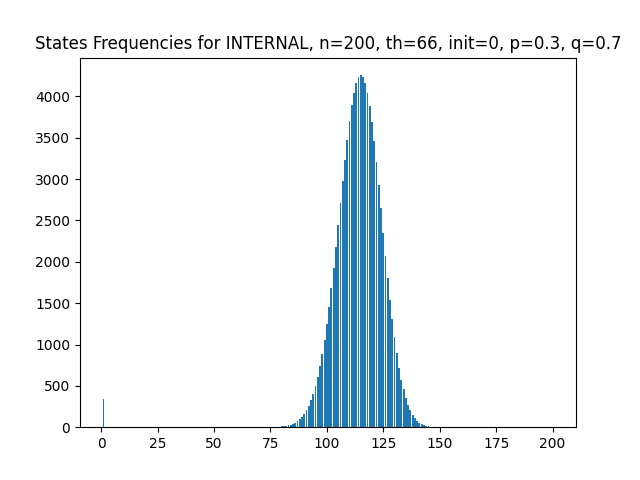}
    \caption{State distribution $p=0.3, q=0.7$}
    \label{fig:internal-distribution:03-07-0}
    \end{subfigure}
    \begin{subfigure}{0.3\textwidth}
    \includegraphics[width=\textwidth]{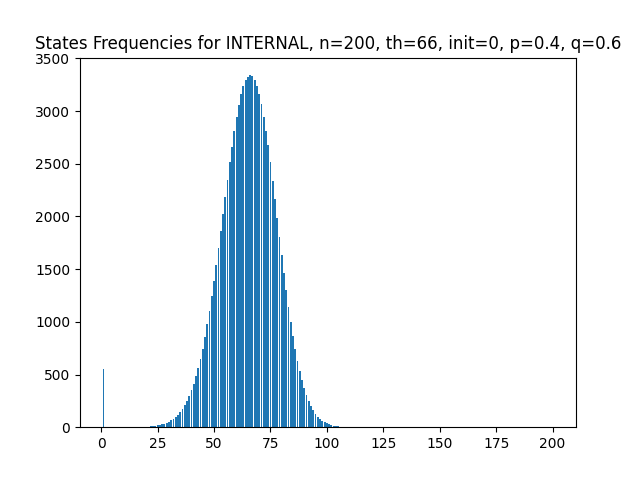}
    \caption{State distribution $p=0.4, q=0.6$}
    \label{fig:internal-distribution:04-06-0}
    \end{subfigure}
    \begin{subfigure}{0.3\textwidth}
    \includegraphics[width=\textwidth]{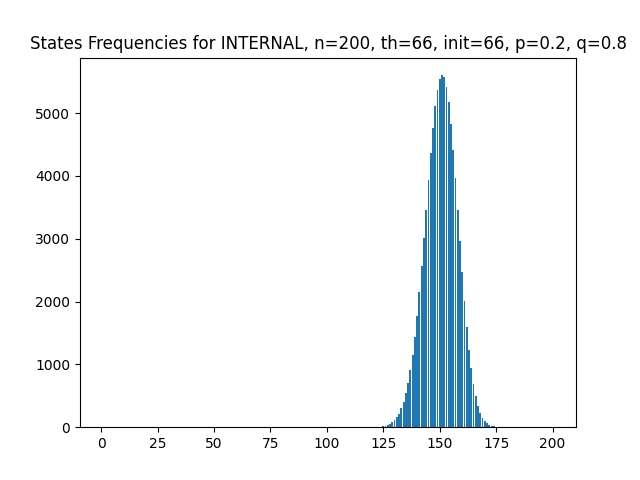}
    \caption{$p=0.2, q=0.8$, init=$66$}
    \label{fig:internal-distribution:02-08-66}
    \end{subfigure}
    \begin{subfigure}{0.3\textwidth}
    \includegraphics[width=\textwidth]{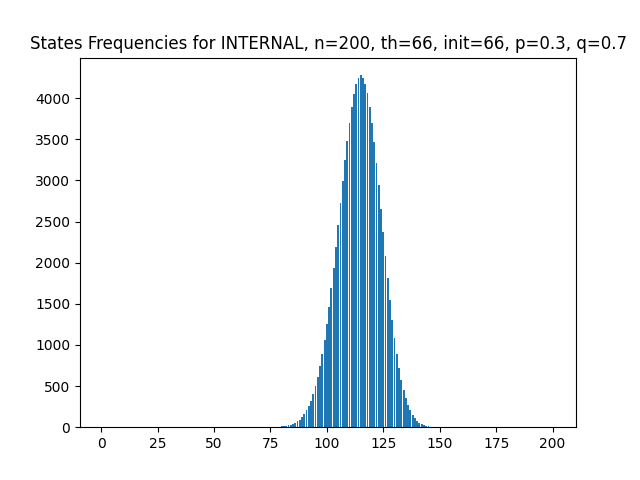}
    \caption{$p=0.3, q=0.7$, init=$66$}
    \label{fig:internal-distribution:03-07-66}
    \end{subfigure}
    \caption{Distribution of time spent at different states when the number of nodes is 200, the initial state is 0 (no Byzantine), and the simulation is run for 100K steps  for the INTERNAL model.}
    \label{fig:internal-distribution}
\end{figure*}

\subsubsection{COORDINATED Results}
\label{sec:ctmc-coordinated-simulations}
The COORDINATED Markovian process is mathematically similar to the EXTERNAL, except that the rate of movement in each state $i$ of the COORDINATED is multiplied by $i$ compared to the EXTERNAL process.
Hence, the distribution of times diagrams are not very interesting in the sense that whenever $p>q$, the Markovian process remains mostly around $0$, even when it is started at $200$.
Similarly, whenever $q>p$, the Markovian process rushes towards $200$, even when it is started at $0$.
Hence, for \ifdefined\CONFversion lack of space, \else brevity, \fi we skip these figures here.

Figure~\ref{fig:COORDINATED-Heatmap-Bad-200-66} illustrates the percentage of totally bad runs, the percentage of runs in which the system flips from bad state to good state, and the average time taken for the system to flip to the first good state in those runs, when the number of nodes is 200, the initial state is 200 (all Byzantine), the threshold is 1/3 (66 nodes), and the simulation is run for 100K time units under the COORDINATED-CTMC model.
As anticipated from the discussion above, it is similar to Figure~\ref{fig:Heatmap-Bad-200-66}.
This similarity applies as well to the other simulation results for this model, dropped for \ifdefined\CONFversion lack of space.\else brevity.\fi


\begin{figure*}[tp!]
    \centering
    \includegraphics[width=\linewidth]{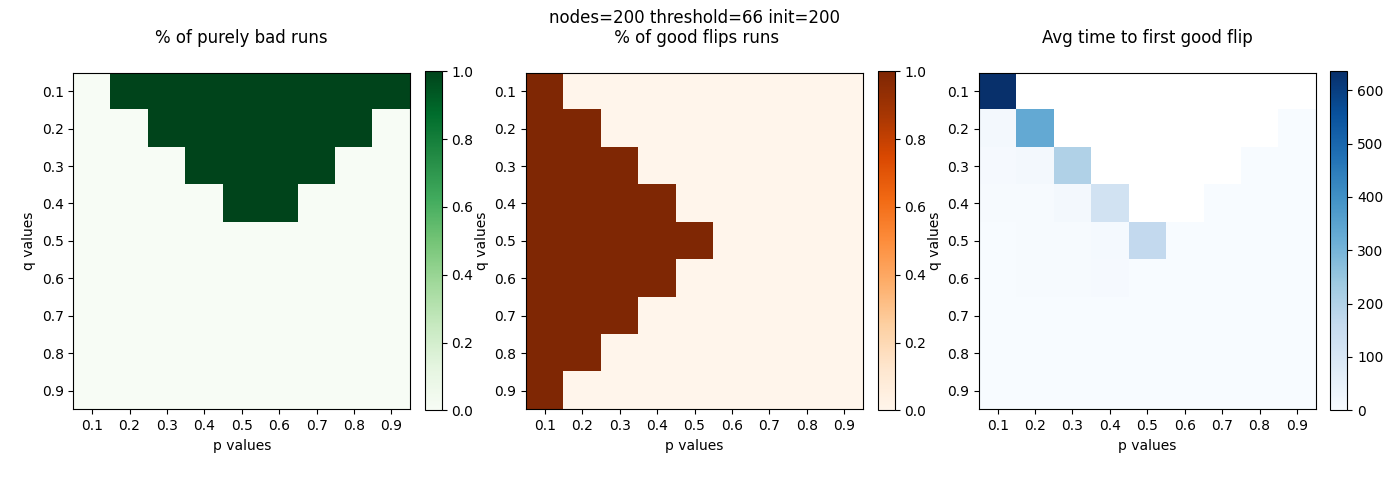}
    \caption{Percentage of totally bad runs, percentage of runs in which the system flips from bad state to good state, and the average time taken for the system to flip to the first good state in those runs; the number of nodes is 200, the initial state is 200 (all Byzantine), the threshold is 1/3 (66 nodes), and the simulation runs for 100K time units in the COORDINATED~model.}
    \label{fig:COORDINATED-Heatmap-Bad-200-66}
\end{figure*}

\section{Conclusion and Future Works}
\label{sec:conclusions}
In this work, we have introduced a new probabilistic variant of the Mobile Byzantine Failures model.
In this model, an attacker may infect correct nodes with a given probability so that they become Byzantine, while the system may detect and cure Byzantine nodes probabilistically as well.
We have explored how to adapt the MAPE-K architecture to this model and applied a formal Markov probabilistic model to analyze the behavior of the system with respect to the (fluctuating) number of Byzantine failures.

In particular, we can analyze how long it would take for the number of Byzantine nodes in the system to reach a given threshold (e.g., a one-third) above which the system is considered unsafe, and must be manually restarted.
Similarly, we can analyze how long it would take a system with too many Byzantine nodes to cure itself to the point where the number of Byzantine nodes falls below the threshold.
If this expected evolution is acceptable, and the system realizes a self-stabilizing Byzantine tolerant protocol~\cite{DD05,DRS23}, we can let the system continue to cure on its own.
Otherwise, a manual reset is required.

We have also simulated the behavior of the system.
These simulation results help visualize the behavior of the system, as backed by the mathematical analysis.
In fact, the visualization of the distribution of time spent in each state, and in particular the non-intuitive behavior discovered for the INTERNAL model led us to perform the mathematical analysis in Section~\ref{sec:state-dist-external}, which gave the mathematical explanation to what we were seeing.
\ifdefined\CONFversion
For lack of space, we could not show all results.
Missing results and simulation code will be open sourced once the anonymity requirement is~lifted.
\fi

\ifdefined\CONFversion
\else
The strategy proposed for the Configuration Planner is only a first strategy to support the automatic control loop that governs the reconfiguration once we approach a state where the distributed system is no longer secure. 
As a future work, we are studying other planning strategies that are able to trade off several cost variables behind the reconfiguration (e.g., trading off- the downtime of the system for the cost of deployment). In addition, we are investigating how to extend the probabilistic model to account for diversity in the~configurations.
\fi

Last, in the classical (fixed) Byzantine model, the longer an instance of a BFT-resilient protocol runs, the higher the probability that it will terminate.
However, in our more generalized mobile Byzantine failure model, depending on the rate parameters $p$ and $q$, as the run becomes longer, the number of Byzantine nodes could cross the threshold required to ensure liveness and safety.
In such cases, the termination probability of classical protocols could drop to below 1, and even safety could become probabilistic with a probability lower than 1.
Analyzing this for specific protocols is left for future work.



\bibliographystyle{IEEEtran}
\bibliography{references,mybibliography}




\end{document}